\documentclass[]{interact}

\usepackage{amsmath}
\usepackage{amssymb}
\usepackage{siunitx,booktabs}
\expandafter\def\csname ver@subfig.sty\endcsname{}
\usepackage{chngpage}
\usepackage{algorithm}
\usepackage[noend]{algpseudocode}
\usepackage{enumitem}
\usepackage{csvsimple}
\usepackage{rotating}
\usepackage{longtable}
\usepackage{multirow}
\usepackage[normalem]{ulem}
\usepackage{picins}
\usepackage{lscape}
\usepackage{booktabs}
\useunder{\uline}{\ul}{}
\newcolumntype{K}[1]{>{\centering\arraybackslash}m{#1}}

\usepackage{epstopdf}

\usepackage[longnamesfirst,sort]{natbib}
\bibpunct[, ]{(}{)}{;}{a}{,}{,}
\renewcommand\bibfont{\fontsize{10}{12}\selectfont}

\begin{document}

\title{Profit-oriented sales forecasting: a comparison of forecasting techniques from a business perspective}

\author{
\name{Tine Van Calster\textsuperscript{a}\thanks{Corresponding author: tine.vancalster@kuleuven.be}, Filip Van den Bossche\textsuperscript{a}, Bart Baesens\textsuperscript{a, b} and Wilfried Lemahieu\textsuperscript{a}}
\affil{\textsuperscript{a} Faculty of Economics and Business, KU Leuven, Naamsestraat 69, 3000 Leuven, Belgium; \textsuperscript{b}University of Southampton, University Road, Southampton SO17 1BJ, United Kingdom}
}

\maketitle

\begin{abstract}
Choosing the technique that is the best at forecasting your data, is a problem that arises in any forecasting application. Decades of research have resulted into an enormous amount of forecasting methods that stem from statistics, econometrics and machine learning (ML), which leads to a very difficult and elaborate choice to make in any forecasting exercise. This paper aims to facilitate this process for high-level tactical sales forecasts by comparing a large array of techniques for 35 times series that consist of both industry data from the Coca-Cola Company and publicly available datasets. However, instead of solely focusing on the accuracy of the resulting forecasts, this paper introduces a novel and completely automated profit-driven approach that takes into account the expected profit that a technique can create during both the model building and evaluation process. The expected profit function that is used for this purpose, is easy to understand and adaptable to any situation by combining forecasting accuracy with business expertise. Furthermore, we examine the added value of ML techniques, the inclusion of external factors and the use of seasonal models in order to ascertain which type of model works best in tactical sales forecasting. Our findings show that simple seasonal time series models consistently outperform other methodologies and that the profit-driven approach can lead to selecting a different forecasting model.
\end{abstract}



\section{Introduction}
\label{Introduction}

This paper focuses on one of the most frequently asked questions in forecasting theory and practice: which technique(s) should I choose to forecast this time series? In literature, this question has been posed many times and has indeed been answered by benchmarks and competitions \citep{a4, c3, p2}, as forecasting has been an integral part of the business decision-making process for decades and is used for this purpose in many industries \citep{a4, c1, l2}. However, most studies only take one evaluation criterion into account, i.e. the performance of the techniques on a test set, while the final choice of a model in a business context depends on more considerations. Undoubtedly, the costs that are associated with inaccurate forecasts make sure that accuracy will always remain an important evaluation standard \citep{k1}. However, from a decision-making perspective, other questions immediately arise in the mind of the business expert as well, such as the potential impact of the forecast on the revenue of the company or the maintenance cost of the model. This paper therefore proposes an expected profit function that can be integrated into several steps of the forecasting process, while also taking a closer look at which types of models perform best in a sales forecast on a tactical or strategic level.

Recent publications have shown a large offering of forecasting techniques, ranging from the statistical methods to machine learning techniques. Given all of these theoretical and technological developments, it is becoming increasingly difficult to select the right type of technique for a given use case. Especially the group of Machine Learning (ML) techniques has received a lot of attention recently, as it constitutes one of the most popular topics in forecasting literature \citep{f2}. Most articles on ML techniques report favourable results when compared to more traditional methodologies, both for single use cases and more extensive comparisons \citep{c3}, although publications generally have a tendency to only report on positive outcomes \citep{a3}. However, several authors have expressed their reservations concerning these complex techniques \citep{m2}. In contrast, \citep{c3} have shown that machine learning has caught up with statistical modelling and should not be dismissed lightly for forecasting exercises. This paper therefore also aims to investigate whether these more complex ML techniques truly outperform the classical models for a tactical sales forecast.
 
In this paper, we will focus on the field of sales forecasting, as successful sales forecasts are vital in both short- and long-term strategic and financial planning \citep{r1}. This research specifically deals with high-level forecasts, which are primarily meant for decision-making purposes, as opposed to inventory planning for specific products. In practice, this typically entails a monthly time series, which is non-intermittent and is prone to display a trend, a seasonal pattern or a combination of these characteristics. This type of time series is common in other fields as well, and has therefore frequently been used for benchmarking purposes \citep{a4, c3, p2}. We therefore take a look at the performance of techniques that model seasonality versus methodologies that do not have this ability, as season is a typical characteristic of sales time series. While trying out non-seasonal models might seems counterintuitive for this data, a lot of the more recently developed techniques do not have a seasonal component and still seem to perform very well for many applications. This paper therefore also investigates whether this type of model can perform well on these seasonal time series, given the necessary pre-processing of the data. Furthermore, this high-level data also raises the question of the usefulness of incorporating external factors into the forecast. While th addition of variables has obvious benefits, such as the explanatory value, it frequently leads to higher model maintainability costs. Thus, we also compare univariate techniques with and without the ability to add external drivers to one another in this paper.

Our contributions are twofold, as we aim to both benchmark a large set of forecasting techniques and integrate a practical construct into the model building and evaluating process, i.e. profit. Firstly, we propose a new strategy to inject a profit-oriented view into the entire forecasting process without explicitly forecasting profit itself. In practice, this constitutes a different way of performing feature selection, tuning hyper parameters and evaluating the forecasting techniques with the goal of achieving the models that yield the highest expected profit. The expected profit function that is used for this purpose, is easy to understand and adaptable to any situation by combining forecasting accuracy with business expertise \citep{v1}. Furthermore, our methodology ensures a completely automated and data-driven model building process. Secondly, we benchmark a large range of forecasting techniques according to three different categorizations. As mentioned above, we contrast a range of complex techniques and traditional techniques, in order to assess whether the ML techniques are truly able to perform equally in regards to tactical sales forecasts. Secondly, we take the seasonal characteristics of sales time series into consideration by distinguishing techniques that model seasonality themselves and methods that require seasonal dummy variables to achieve the same goal. Finally, we contrast techniques with and without variables, as we investigate the value of external factors in a high-level sales forecast. In terms of evaluating the techniques, we take accuracy, expected profit, model complexity and model interpretability into consideration in order to integrate the business aspect of forecasting into the benchmark. In the end, we aim to quantitatively select the techniques that forecast accurately, lead to the highest expected profit for any business case, and make the most sense from a business perspective. We will address these research questions by means of a total of 35 monthly sales datasets. The datasets were collected from both The Coca-Cola Company and from publicly available resources in order to add to the generalizability of the study.

The paper is organized as follows. Section \ref{Rel} deals with the related work that provides a necessary background to the research questions. Section \ref{meth} describes the datasets, the forecasting techniques and the general methodology of the experiments. Next, section \ref{Res} focuses on the results of the research, while section \ref{Conc} includes the conclusion.

\section{Related work}
\label{Rel}

This section on related work focuses on the necessary background literature for the research questions. We take a closer look at the forecasting literature on benchmarking, while also considering recent literature on profit-oriented analytics.

\subsection{Benchmarking in forecasting}
\label{benchfor}

Forecasts are typically performed by three categories of techniques \citep{c1}: traditional time series analysis \citep{a1, a2, a5, a6, f3, g1, g2, p2, r1, s2}, causal regression techniques \citep{a2, a5, b4, c1, l1, m1, n2}, and more complex artificial intelligence techniques \citep{a2, b4, c1, c3, f1, l1, t2}. The emergence of new techniques often requires a comparison with former methods, which leads to an extensive literature on benchmarking, both for individual use cases \citep{a1, a5, b4, g1, g2, l1} and for larger sets of time series \citep{a6, c1, c3, f3, m1, m2, p2, w1}. This research consists of both field-specific \citep{a1,a2, a6, b4, c1, f1, l1, m1, w1} and industry-neutral benchmarks, which are oriented towards general conclusions \citep{c3, m2, p2}. While some studies use a combination of generated data and industry data \citep{p2}, most use real-life datasets to answer their research questions \citep{b4, c1, l1, w1}.

In terms of the conclusions that have come out of the larger studies, some discrepancies arise. While several studies point out that the newer ML techniques do not perform as well as the more traditional methods for classical time series \citep{m2}, others claim that these complex techniques have caught up in recent years \citep{c3}. In this paper, we therefore take a look at a wider range of techniques from all three categories that were mentioned above. Furthermore, we also contrast techniques with and without external factors, which adds another factor that has not been part of many larger benchmarking studies, except for \citep{a6}. Our paper combines these elements in an extensive benchmark that is based on publicly available data and recent sales time series.

\subsection{Profit-driven analytics}
\label{profan}

Profit-driven analytics has recently become a hot topic in analytics, as businesses are interested in the actual value that predictive models generate or the influence that they have on their eventual net profits. Integrating this value-centric view into analytics, has led to a growing number of profit-driven methodologies, techniques and metrics \citep{v3}. These profit functions can be used in different steps of the model building and model selection process. For example, profit has been used as an evaluation metric for benchmarks in different fields \citep{o2, v4}, while it has inspired entire profit-driven algorithms as well \citep{s4, v2}. In this paper, we aim to integrate this profit-oriented view into multiple steps of the forecasting process instead of only using it as an additional evaluation criterion.

In forecasting, research on the profit aspect is scarcer than in other fields. While the monetary value of classification models has been extensively reviewed, the same cannot be said for regression models. However, the impact of forecasting accuracy on net profit is an interesting subject, as under- and over-forecasting both lead to completely different costs. The former might lead to a loss in sales and out-of-stock products, while the latter can lead to overstock and storage costs. While both directions for the error inevitably bring about a loss of profit, they are often not equal. Completely symmetric profit loss functions that are solely based on accuracy measures are therefore not representative of the real world. The ultimate goal of profit-oriented analytics is to find the model with the best balance between costs and accuracy. While these two concepts are inevitably linked in a forecasting exercise, we cannot state that they are exactly the same. Therefore, profit-oriented benchmarking should take into account both traditional accuracy metrics and metrics that point to the costs of the forecast, such as expected profit functions or model complexity. So far, two different views on the integration of profit into forecasting exercises have been proposed in recent literature. The first perspective optimizes an asymmetric loss function during model training to model the imbalance between over- and under-forecasting. \citep{c4} applies this methodology to neural networks, while \citep{y1} take a closer look at support vector regression models. The second way of integrating profit into a forecasting function takes place after the training process. \citep{b1} propose a tuning procedure that modifies the predictions so they are cost-optimal. \citep{z2} further fine-tune this procedure. \citep{b4} also take a monetary value into account when evaluating their forecasts, but do not modify the models in any way.

In this paper, we take profit into account in all of the steps that are mentioned above. We optimize the parameters of our models, select features when necessary and evaluate our forecasts based on an asymmetric expected profit function that can easily be adjusted to any business case.

\section{Methodology}
\label{meth}

This methodology section is divided into five parts. We begin by describing the datasets and by explaining the profit function that was used for both optimization and evaluation purposes in this paper. Next, the general experimental set-up is introduced, which also includes the description of the feature selection procedure. The fourth subsection is dedicated to an overview of the forecasting techniques, while the last subsection focuses on evaluation metrics.

\subsection{Data}
\label{data}

The data sets in this paper stem from two sources. Firstly, The Coca-Cola Company has given us a total of 20 time series, which represent two of their product categories in ten different countries. These monthly time series all range from January 2004 until September 2016. The external variables that correspond with these datasets, were collected by means of in-company data sources and are all based on information about the location of the data. Concretely, they consist of 20 variables that contain information on weather, macro-economic indicators, holidays and pricing information. As weather information, 4 variables were included, such as temperature and precipitation, while 9 variables allude to macro-economic information, such as GDP and CPI. Additionally, 3 factors refer to calendar effects of public holidays, while the final 4 variables relate to both in-company and competitor pricing. An overview of these variables can be found in Table \ref{tab:var}. These external factors were selected according to data availability, but also take into account the literature on the interesting types of variables for sales forecasting. Several types of information have proven to be useful in this field, although this generally depends on the aggregation level of the time series \citep{s5} and the volatility of the time series \citep{c5}. Research has shown that factors such as weather \citep{b2}, macro-economic influences \citep{s1} and pricing and promotional information \citep{h1, m1} all have an impact on sales.

\begin{table}[]
\centering
\resizebox{\textwidth}{!}{%
\begin{tabular}{@{}ll@{}}
\textit{\textbf{Variable name}}      & \textit{\textbf{Explanation}}                                                                     \\ \midrule
\textit{Weather}                     &                                                                                                   \\ \midrule
\textbf{Maximum temperature}         & Average daily maximum temperature weighted by population                                          \\
\textbf{Maximum temperature squared} & Square of average daily maximum temperature weighted by population                                \\
\textbf{Precipitation}               & Average daily precipitation volume                                                                \\
\textbf{Sunshine hours}              & Average daily number of sunshine hours                                                            \\ \midrule
\textit{Macro-economic indicators}   &                                                                                                   \\ \midrule
\textbf{Consumer Price Index}        & Seasonally adjusted percentage change of CPI with regards to the previous month                   \\
\textbf{Unemployment rate}           & Percentage of unemployment for entire population                                                  \\
\textbf{Exchange rate}               & Exchange rate with US dollar                                                                      \\
\textbf{Short-term interest rate}    & Short-term interest rate in percentage per annum                                                  \\
\textbf{Industrial production}       & Seasonally adjusted percentage change of industrial production with regards to the previous month \\
\textbf{Merchandise import}          & Seasonally adjusted percentage change of Merchandise import with regards to the previous month    \\
\textbf{Merchandise export}          & Seasonally adjusted percentage change of Merchandise export with regards to the previous month    \\
\textbf{Gross Domestic Product}      & Seasonally adjusted annual rate, percentage change of GDP with regards to the previous month      \\
\textbf{Private Consumption}         & Seasonally adjusted annual rate, percentage change of PC with regards to the previous month       \\ \midrule
\textit{Holidays}                    &                                                                                                   \\ \midrule
\textbf{Public holiday}              & Number of public holidays per month                                                               \\
\textbf{Weekend}                     & Number of public holidays in the weekend per month (possibility of long weekend)                  \\
\textbf{Tuesday/Thursday}            & Number of public holidays on Tuesday or Thursday per month (possibility of long weekend)          \\ \midrule
\textit{Pricing}                     &                                                                                                   \\ \midrule
\textbf{Company price}               & Average product category price in US dollars                                                      \\
\textbf{Company price deflated}      & Average product category price in US dollars deflated by CPI                                      \\
\textbf{Competitor price}            & Average product category price of the main competitor in US dollars                               \\
\textbf{Competitor price deflated}   & Average product category price of the main competitor in US dollars deflated by CPI 
\\ \bottomrule             
\end{tabular}%
}
\caption{Summary of external variables}
\label{tab:var}
\end{table}

Secondly, we include a total of 15 publicly available datasets with similar characteristics in the analyses, in order to increase the generalizability of our findings, which can mostly be found in The Time Series Data Library\footnote{\url{https://datamarket.com/data/list/?q=provider:tsdl}}\textsuperscript{,}\footnote{\url{https://opendata.socrata.com/Business/Car-Sales-Data/da8m-smts}}. The general features of these monthly time series are summarized in Table \ref{tab:pubdata}. As all of these datasets also include information on location, we collected twelve external variables that contain information on weather, macro-economic indicators and holidays as well. Concretely, we include four weather variables, seven macro-economic indicators and one holiday variable. The weather variables consist of the same features as defined in Table \ref{tab:var}, while the macro-economic information includes all features in Table \ref{tab:var} except Merchandise Import and Merchandise Export. Finally, the models with external factors also contain the number of public holidays for each month. Pricing information was not available for these datasets. 
The sources for these three categories are publicly available\footnote{\url{https://crudata.uea.ac.uk/cru/data/hrg/cru_ts_3.23/crucy.1506241137.v3.23/}}\textsuperscript{,}\footnote{\url{https://data.oecd.org/}}\textsuperscript{,}\footnote{\url{https://pypi.python.org/pypi/holidays}}.

\begin{table}[]
\centering
\resizebox{\textwidth}{!}{%
\begin{tabular}{@{}lclcl@{}}
\textbf{Name}        & \textbf{\begin{tabular}[c]{@{}c@{}}Number of \\ product categories\end{tabular}} & \textbf{Range}                & \textbf{\begin{tabular}[c]{@{}c@{}}Number of \\ data points\end{tabular}} & \textbf{Location} \\ \midrule
\textbf{Beer}        & 1                                                                                & January 1956 -– August 1995    & 476                                                                       & Australia         \\
\textbf{Car sales 1} & 1                                                                                & January 1996 -– December 2008  & 156                                                                       & California        \\
\textbf{Car sales 2} & 1                                                                                & January 1960 -– December 1968  & 108                                                                       & Canada            \\
\textbf{Champagne}   & 1                                                                                & January 1964 -– September 1972 & 105                                                                       & France            \\
\textbf{Paper}       & 1                                                                                & January 1963 -– December 1972  & 120                                                                       & France            \\
\textbf{Petrol}      & 4                                                                                & January 1971 -– December 1991  & 252                                                                       & USA               \\
\textbf{Wine}        & 6                                                                                & January 1980 -– July 1995      & 187                                                                       & Australia 
\\ \bottomrule       
\end{tabular}%
}
\caption{Public data summary}
\label{tab:pubdata}
\end{table}

\subsection{Expected profit function}
\label{exprof}

The evaluation of any predictive model is generally focused on the accuracy that it achieves on a test set. In this paper, however, we take both accuracy and a more business-oriented profit measure into account. The profit measure is represented by Equation \ref{eq2}, which is dependent on our definition of the Percentage Error (PE), which can be found in Equation \ref{eq1}. This profit measure was first defined in \citep{v1} and represents an estimation of the expected profit of the target variable. The formula is very easy to interpret and can easily be adjusted to any business use case. The two fundamental components of the expected profit measure are the volume of the sales, as more sales lead to more profit, and the accuracy of the forecast, as bad forecasts inevitably lead to a loss of profit. Next to these two core elements, we introduce several parameters that integrate expert knowledge into the profit function.

\bigskip
\noindent
\begin{equation}
\label{eq1}
PE=\frac{Actuals_i - Forecast_i}{Actuals_i}*100
\end{equation}
\indent

\noindent
\begin{adjustwidth}{-1cm}{-1cm}
\begin{equation}
\label{eq2}
Profit = 
  \begin{cases} 
   ((1-(\alpha*|PE|))*(\beta_{cat}*Volume_{cat}) &     PE > \gamma \text{ or } PE < \delta \\
   \beta_{cat}*Volume_{cat}						 & otherwise
  \end{cases}
\end{equation}
\end{adjustwidth}
\indent
\bigskip

Firstly, the business user can influence the impact of the forecasting error on the expected profit by setting two parameters. The first one deals with how the size of the error is used as a penalization, as both over- and under-forecasting have proven to lead to various costs \citep{k1}. This penalization factor $\alpha$ can be modified according to a specific circumstance in a data-driven manner by executing a sensitivity analysis on a validation set. In this instance, α is set at $1.5\%$, which was determined in \citep{v1}. Furthermore, the business expert can set penalization boundaries $\gamma$ and $\delta$, which indicate that any forecast that has a PE within these boundaries, does not lead to a significant impact on the final profit. Note that $\gamma$ should always be larger than $\delta$. For example, we set boundaries of $1\%$ error in both directions for The Coca-Cola Company use case ($\gamma = 1$ and $\delta = -1$). This leads to a preference for models with a small over-forecast, as the expected profit is higher because of a larger volume of sales in that case. This tendency is deemed appropriate for the use case because the company considers this an investment in the future. However, the $\gamma$ and $\delta$ parameters can also be set unequally, if the forecasting error has a larger impact on profit in one particular direction, or even be completely omitted, if every inaccuracy while forecasting leads to a loss of profit.

Secondly, the $\beta_{cat}$ weight refers to the profit margin for the product or product category at hand. This weight can be expressed both relatively between different products and in absolute numbers, such as currencies. For The Coca-Cola Company use case, these $\beta$ weights were determined by the profit that the product actually generated in the last year of the original training set. It is important to note that these weights remain constant throughout the analyses once they are set by the training set of the first prediction. The actual profit of a product will fluctuate over time and is driven by many external factors that are not captured in the function. We have chosen to keep this parameter constant because of two reasons: ease of use and availability of profit data. While the first reason is self-explanatory, the second one is tied to the particular use case of this paper. If data about the actual profit of a product is more readily available, this parameter can be used dynamically by updating it during the testing process. The profit in the analyses of this paper can therefore be viewed as the profit that the product will generate if business stays the same and must truly be interpreted as the expected profit. The $\beta$ weights for the publicly available datasets were chosen randomly with values between 0 and 3, and are displayed in Table \ref{tab:weights}.

\begin{table}[]
\centering
\begin{tabular}{@{}lr@{}}
\toprule
\textit{\textbf{Name}} & \multicolumn{1}{l}{\textit{\textbf{$\beta$ weights}}} \\ \midrule
\textbf{Beer}          & 0.1                                             \\
\textbf{Car sales 1}   & 2.1                                             \\
\textbf{Car sales 2}   & 0.6                                             \\
\textbf{Champagne}     & 1.0                                             \\
\textbf{Paper}         & 1.9                                             \\
\textbf{Petrol}        & 0.1, 1.2, 0.1, 2.8                              \\
\textbf{Wine}          & 2.2, 2.2, 2.1, 1.5, 0.4, 2.7                    \\ \bottomrule
\end{tabular}
\caption{$\beta$ weights of public datasets}
\label{tab:weights}
\end{table}

\subsection{Experimental set-up}
\label{expset}

\begin{figure}[h]
    \centering
\includegraphics[width=13cm, clip]{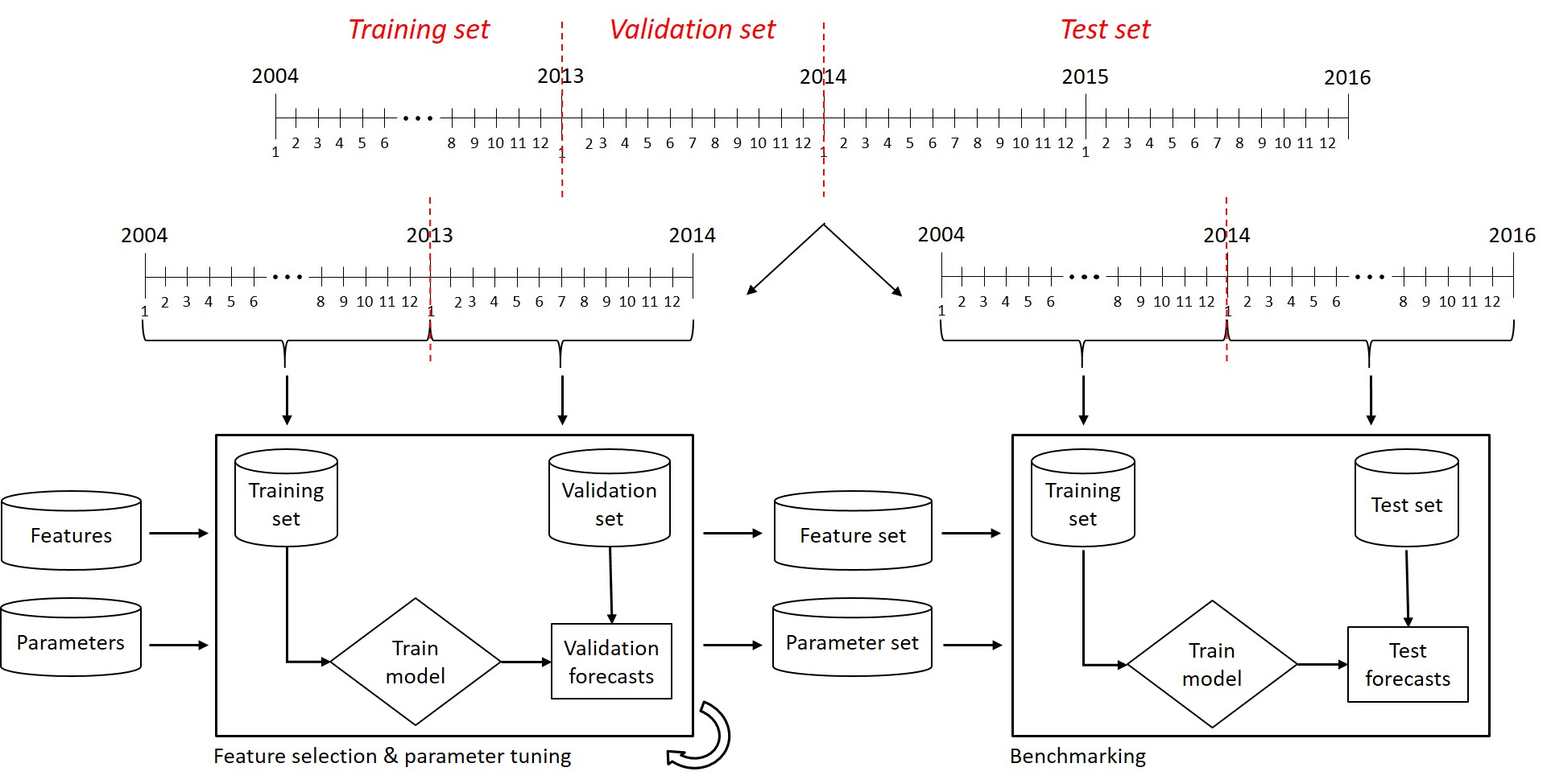}
 \caption{Experimental set-up}
    \label{fig:setup}
\end{figure}

The general experimental set-up consists of hold-out sample forecasts for all datasets. Concretely, the time series are split up into training, validation and test sets. The test set includes the final two years of the data, which leads to 24 data points to forecast. The validation set then consists of the year before the date that will be forecast and is only used for feature selection and parameter tuning when necessary for the given technique. Parameter tuning is performed once on the first validation set in the testing procedure, in order to avoid computational issues in the testing procedure. However, the feature selection procedure is repeated every three months, in order to keep the model up-to-date. Once the necessary variables and hyper parameters have been selected, the training and validation sets are merged together in order to forecast the test set. Both the training and validation sets change with every forecast, as the set-up consists of an expanding window. In the end, we therefore collect 24 one-month ahead forecasts for each technique and for each dataset. The complete experimental set-up is visualized in Figure \ref{fig:setup}.

The feature selection procedure consists of a hybrid method, which is based on the combination of Minimum Redundancy Maximum Relevance criterion (mRMR) that was created by \citep{p1} as a filtering technique, and a simple incremental wrapper method. The mRMR method is a mutual-information based algorithm that ranks the external factors according to their shared information with the target variable, while also taking into account their dependency on the other external factors. This can be achieved by finding the feature set $S$ with $m$ features $x_i$ that maximizes the relevance with the target class $c$ and minimizes the dependency between the independent variables. In short, this filter finds the features that maximize Equation \ref{eq3}, which combines Equation \ref{eq4}  (Dependency) and Equation \ref{eq5} (Redundancy).

\bigskip
\bigskip
\noindent
\begin{equation}
\label{eq3}
max \phi(D,R), \phi= D - R
\end{equation}
\indent

\noindent
\begin{equation}
\label{eq4}
D=\frac{1}{| S |}\sum_{\forall x_i \in S}I(x_i , c)
\end{equation}
\indent

\noindent
\begin{equation}
\label{eq5}
R=\frac{1}{| S |^2}\sum_{\forall x_i , x_j \in S}I(x_i , x_j)
\end{equation}
\indent

In this paper, the first feature step in this paper selects either the top 15 or the top 10 ranking of features, for the Coca-Cola Company datasets and the public datasets respectively, and then passes this on to the next step. Next, a simple forward incremental wrapper method starts with the top feature of the ranking and forecasts the validation set. Consecutively, one feature is added at a time into the feature set until the entire top 15 or top 10 ranking is used in the forecasting model. This methodology therefore takes advantage of the initial ranking that was made by the mRMR filter. The feature set that will be used to forecast the test set, is selected out of these 15 or 10 options by maximizing the profit function, which is defined in Section \ref{exprof}. This entire procedure is explained in Algorithm \ref{pseudo}. In our benchmark, $k$ is either 15 or 10, depending on the dataset at hand, and $m$ is equal to 12 months.

\begin{algorithm}[H]
\caption{Pseudo code for feature selection procedure}\label{pseudo}
\begin{algorithmic}
\State choose size of validation set $m$
\State split time series into training set $S_{tr}$ and validation set $S_{val,m}$
\State choose initial number of features $k$
\State rank features according to the mRMR Maximum Relevance criterion into ranking $R_k$
\For {$i = 1$ to $k$}
\State select top $i$ features from $R_k$
\For {$j = 1$ to $m$}
\State train model with $R_i$ features on training set $S_{tr}$
\State forecast $S_{val,j}$
\State calculate profit $P_{i,j}$
\State add $S_{val,j}$ to training set $S_{tr}$
\EndFor
\State \textbf{end for}
\State calculate profit $P_i$ by summing over all $P_{i,m}$
\State reset training set $S_{tr}$ and validation set $S_{val,m}$ to original split
\EndFor
\State \textbf{end for}
\State Select $R_i$ features with highest profit $P_i$
\end{algorithmic}
\end{algorithm}

Feature selection is generally important because of two entirely different reasons. Firstly, some of the variables might be correlated or influenced by the same underlying information, which can lead to less accurate forecasts \citep{b3}. A feature selection procedure is therefore used to determine which set of variables has the highest predictive power, while also eliminating any possible multicollinearity. Secondly, feature selection is equally important from a business perspective, as transparent models also have an explanatory advantage. Business analysts are interested in gaining knowledge on which external factors might influence their target variable, which can be useful for strategic decisions \citep{a6}. However, this knowledge then also relates to the maintenance of the model, as the variables that never survive the feature selection procedure during testing, are not needed any longer.

\subsection{Forecasting techniques}
\label{tech}

In order to conduct the necessary experiments, a total of 17 forecasting techniques were selected, which are summarized in Table \ref{tab:tech}. These techniques are categorized according to three different types of attributes in order to answer our research questions. Firstly, we organize the methods according to the ability to use them as univariate with and/or without external drivers. We define a univariate technique without variables as a technique that only makes use of the sales times series itself to predict the next month. Techniques that are able to include variables, however, also integrate the external drivers, such as the weather, to generate a prediction. 8 techniques can be used in both ways, such as regression models, when past sales values are encoded as independent variables, next to the aforementioned external factors. We therefore benchmark a total of 26 techniques in our final analysis. Secondly, Table \ref{tab:tech} displays the ability of a technique to explicitly model the seasonality of a time series, as seasonality is a typical characteristic of the sales time series that we are considering in this paper. Thirdly, the forecasting techniques are classified into Machine Learning (ML) techniques and non-ML techniques. Recently, a lot of forecasting literature has focused on these ML techniques and often reports them to be more accurate than traditional techniques. In order to simplify the issue of what is considered an ML technique and what is not, we chose to consider methods ML if they belong to one of the four following categories: decision tree learning, neural networks, support vector machines and k-nearest neighbours, as this last category is based on a clustering algorithm. These three categorizations will underpin the answer to which type of technique is best used to achieve an accurate sales forecast. Finally, the table also contains the hyper parameters that were selected beforehand, and their possible values. Tuning hyper parameters has proven to be essential for truly assessing how well a certain technique can perform \citep{c2}, and is therefore an essential part of benchmarking in general. In this paper, the parameter selection was conducted by evaluating model performance on the validation set and by applying an exhaustive grid search methodology. The evaluation metric that was optimized, is again the expected profit function that was defined in Section \ref{exprof}. Note that only the parameters that are mentioned in \ref{tab:tech} are set in this way.

\begin{table}[]
\centering
\resizebox{\textwidth}{!}{%
\begin{tabular}{@{}llllll@{}}
\toprule
\textbf{Model}                                                                                & \textbf{Variables?} & \textbf{Seasonal?} & \textbf{ML?} & \textbf{Hyper parameters}                                                                                                   & \textbf{Possible values}                                                                                             \\ \midrule
\textbf{\begin{tabular}[c]{@{}l@{}}Holt-Winters exponential\\   smoothing\end{tabular}}       & No           & Yes                & No           & /                                                                                                                           & /                                                                                                                    \\
\textbf{Seasonal ARIMA}                                                                       & No           & Yes                & No           & \begin{tabular}[c]{@{}l@{}}AR, MA, SAR and\\ SMA terms\end{tabular}                                                         & {[}0, 5{]}                                                                                                           \\
\textbf{\begin{tabular}[c]{@{}l@{}}Seasonal decomposition by\\   Loess model\end{tabular}}    & No           & Yes                & No           & /                                                                                                                           & /                                                                                                                    \\
\textbf{Seasonal random walk}                                                                 & No           & Yes                & No           & /                                                                                                                           & /                                                                                                                    \\
\textbf{ARMA-GARCH}                                                                           & No           & No                 & No           & AR and MA terms                                                                                                             & {[}0, 5{]}                                                                                                           \\
\textbf{Random walk}                                                                          & No           & No                 & No           & /                                                                                                                           & /                                                                                                                    \\ \midrule
\textbf{Seasonal ARIMAX}                                                                      & Yes         & Yes                & No           & \begin{tabular}[c]{@{}l@{}}AR, MA, SAR and\\ SMA terms\end{tabular}                                                         & {[}0, 5{]}                                                                                                           \\
\textbf{Vector Autoregression}                                                                & Yes         & No                 & No           & AR term                                                                                                                     & {[}0, 5{]}                                                                                                           \\ \midrule
\textbf{\begin{tabular}[c]{@{}l@{}}Conditional Inference\\   Regression Tree\end{tabular}}    & Both                 & No                 & Yes          & /                                                                                                                           & /                                                                                                                    \\
\textbf{Multiple Linear Regression}                                                           & Both                 & No                 & No           & /                                                                                                                           & /                                                                                                                    \\
\textbf{\begin{tabular}[c]{@{}l@{}}Multivariate Adaptive\\   Regression Splines\end{tabular}} & Both                 & No                 & No           & \begin{tabular}[c]{@{}l@{}}Maximum degree of \\ interaction\end{tabular}                                                    & {[}1, 2{]}                                                                                                           \\
\textbf{\begin{tabular}[c]{@{}l@{}}Recursive Partitioning\\   Regression Tree\end{tabular}}   & Both                 & No                 & Yes          & /                                                                                                                           & /                                                                                                                    \\
\textbf{K Nearest Neighbors Regression}                                                       & Both                 & No                 & Yes          & \begin{tabular}[c]{@{}l@{}}Number of neighbors\\   Weights for neighboring \\ response values\end{tabular}                  & \begin{tabular}[c]{@{}l@{}}{[}2, 5{]}\\   uniform, by distance\end{tabular}                                          \\
\textbf{Long Short Term Memory RNN}                                                           & Both                 & No                 & Yes          & Number of hidden neurons                                                                                                    & {[}1, 10{]}                                                                                                          \\
\textbf{Random Forests}                                                                        & Both                 & No                 & Yes          & /                                                                                                                           & /                                                                                                                    \\
\textbf{Simple Multilayer Perceptron}                                                         & Both                 & No                 & Yes          & Number of hidden neurons                                                                                                    & {[}1, 10{]}                                                                                                          \\
\textbf{Support Vector Regression}                                                            & Both                 & No                 & Yes          & \begin{tabular}[c]{@{}l@{}}Kernel\\ \\   Penalty parameter \\ of error term \\   Gamma (for rbf kernel\\ only)\end{tabular} & \begin{tabular}[c]{@{}l@{}}Radial basis function,\\ linear\\   1e0, 1e1, 1e2,1e3\\ \\   {[}1e-2, 1e2{]}\end{tabular} \\ \bottomrule
\end{tabular}%
}
\caption{Overview of forecasting techniques}
\label{tab:tech}
\end{table}

It is important to comment on the influence of the type of technique on the data preprocessing aspect of the analyses. Firstly, we normalized all variables to a range between 0 and 1 for all of the analyses in this paper. This step was especially necessary for techniques such as neural networks, as literature reports this as a general practice because they benefit greatly from this step \citep{s3}. Furthermore, business users can derive insights on the relative importance of variables if the forecasting technique is transparent, in order to identify the most important drivers of their sales. Secondly, the time series that are part of the analyses all display a certain trend and seasonality, which should be incorporated into the forecasting model if possible. The time series analysis techniques that we consider in this paper, explicitly include this seasonality in their model building by, for example, defining seasonal parameters. However, other types of techniques, such as regression models or neural networks, do not have this ability, which can lead to worse forecasts if the trend and season have a strong influence on the sales \citep{z1}. We therefore add two additional data preprocessing steps for this type of models: trend/seasonal differencing and seasonal dummy variables. In the first step, we check whether the time series actually contains either a trend or a season by means of appropriate unit root tests, such i.e. the Augmented Dickey-Fuller test \citep{d1} and the Osborn-Chui-Smith-Birchenhall test respectively \citep{o1}. If the results thereof show signs of either characteristic, we apply the corresponding differencing. Secondly, if the time series is seasonal, we also include a set of seasonal dummy variables to further model the possible seasonal effects. These variables are not included in the feature selection procedure, but are always included if there is a seasonal component in the time series. Thirdly, when techniques can be used both with and without variables, past sales values need to be encoded as independent variables. We therefore need to determine how many past values will be included into the model. This hyper parameter is selected on the same validation set as the other hyper parameters, and has possible values ranging from one month to seven months. Furthermore, we define methods with external factors as techniques that use both past sales data and external parameters as independent variables. In this case, the number of past months to use as input to the model, is therefore again a hyper parameter.

Finally, we note that the list of forecasting techniques is not exhaustive. Two types of methods are notably under-represented: ensemble methodologies and deep learning methods. We opted to include only one technique of each category in order to keep the scope of the paper manageable, i.e. Random Forests and Long-Short Term Memory Neural Networks respectively. However, the obvious next step of this research is to take a closer look at these types of methodologies.

\subsection{Evaluation}
\label{eval}

Evaluation for forecasting benchmarks is often entirely based on accuracy metrics. There has been a lot of discussion in the past about which metric gives the best overview of performance when comparing techniques, as many commonly used measures can exhibit strange behavior \citep{h2, k2, t1}. In this paper, we therefore propose a combination of frequently used accuracy metrics and the expected profit function that was defined above, to select the best-performing models. In the first category, we take into account the Mean Absolute Percentage Error (MAPE) and the Root Mean Squared Error (RMSE), as defined in Equations \ref{eq6} and \ref{eq7}. Furthermore, we include the seasonal version of the Mean Absolute Scaled Error, which was first defined in \citep{h2}, based on the seasonality of the time series data. The formula for this metric can be found in Equation \ref{eq8} with $m$ as the seasonality of the time series.  This metric compares a technique's performance to the in-sample error of a seasonal naïve model, which makes it perfect for truly benchmarking techniques. Next to the expected profit function, we also consider the computation time of each forecast as an approximate of the model complexity. We therefore include a total of five quantitative performance metrics in our analysis.

\bigskip
\noindent
\begin{equation}
\label{eq6}
MAPE=\frac{1}{n}\displaystyle\sum_{t=1}^{n}|\frac{Actual_t-Forecast_t}{Actual_t}|*100
\end{equation}
\indent

\noindent
\begin{equation}
\label{eq7}
RMSE=\sqrt{\frac{1}{n}\displaystyle\sum_{t=1}^{n}(Actual_t-Forecast_t)^2}
\end{equation}
\indent

\noindent
\begin{equation}
\label{eq8}
MASE=\frac{\displaystyle\sum_{t=1}^{T}|Actual_t-Forecast_t|}{\frac{1}{T-m}\displaystyle\sum_{t=m+1}^{T}|Actual_t-Actual_{t-m}|}
\end{equation}
\indent

\section{Results}
\label{Res}

The result section of this paper will firstly take a look at the experimental results, which are based on forecasting the 35 datasets with 17 different forecasting techniques. Secondly, we will discuss the implications of these results, while we also comment on the limitations of this study.

\subsection{Experimental results}
\label{Expres}

\begin{table}[]
\centering
\resizebox{\textwidth}{!}{%
\begin{tabular}{@{}rrrrrr@{}}
\toprule
\multicolumn{1}{l}{\textbf{Model}}                            & \multicolumn{1}{c}{\textbf{MAPE}} & \multicolumn{1}{c}{\textbf{RMSE}} & \multicolumn{1}{c}{\textbf{MASE}} & \multicolumn{1}{c}{\textbf{Profit}} & \multicolumn{1}{c}{\textbf{Time}} \\ \midrule
\multicolumn{1}{l}{\textit{Without external factors}}                       & \multicolumn{1}{l}{}              & \multicolumn{1}{l}{}              & \multicolumn{1}{l}{}              & \multicolumn{1}{l}{}                & \multicolumn{1}{l}{}                      \\ \midrule
\textbf{ARMA-GARCH (GARCH)}                                   & 12.16 (\underline{0.00})                      & 12.14 (\underline{0.00})                      & 12.16 (\underline{0.00})                      & 12.58 (\underline{0.00})                        & 19.68 (\underline{0.00})                              \\
\textbf{Conditional Inference Regression Tree (CtreeUni)}     & 12.93 (\underline{0.00})                      & 12.92 (\underline{0.00})                      & 12.88 (\underline{0.00})                      & 12.79 (\underline{0.00})                        & 8.16 (\underline{0.00})                               \\
\textbf{Holt-Winters exponential smoothing (HW)}              & 10.52 (1.00)                      & 10.52 (1.00)                      & 10.54 (1.00)                      & 10.74 (\underline{0.00})                        & 5.61 (\underline{0.00})                               \\
\textbf{K Nearest Neighbors Regression (KNNUni)}              & 13.10 (\underline{0.00})                      & 13.10 (\underline{0.00})                      & 13.12 (\underline{0.00})                      & 12.95 (\underline{0.00})                        & 17.35 (\underline{0.00})                              \\
\textbf{Long Short Term Memory RNN (LSTMUni)}                 & 17.23 (\underline{0.00})                      & 17.26 (\underline{0.00})                      & 17.27 (\underline{0.00})                      & 16.78 (\underline{0.00})                        & 25.05 (\underline{0.00})                              \\
\textbf{Multiple Linear Regression (LRUni)}                   & 13.49 (\underline{0.00})                      & 13.49 (\underline{0.00})                      & 13.46 (\underline{0.00})                      & 13.93 (\underline{0.00})                        & 3.50 (0.61)                               \\
\textbf{Multivariate Adaptive Regression Splines (MARSUni)}   & 13.48 (\underline{0.00})                      & 13.48 (\underline{0.00})                      & 13.46 (\underline{0.00})                      & 13.70 (\underline{0.00})                        & 6.68 (\underline{0.00})                               \\
\textbf{Random Forests (RFUni)}                                & 14.07 (\underline{0.00})                      & 14.07 (\underline{0.00})                      & 14.05 (\underline{0.00})                      & 13.91 (\underline{0.00})                        & 12.16 (\underline{0.00})                              \\
\textbf{Random walk (RW)}                                     & 18.43 (\underline{0.00})                      & 18.41 (\underline{0.00})                      & 18.42 (\underline{0.00})                      & 18.11 (\underline{0.00})                        & \textbf{2.52 (/)}                         \\
\textbf{Recursive Partitioning Regression Tree (RpartUni)}    & 13.58 (\underline{0.00})                      & 13.59 (\underline{0.00})                      & 13.57 (\underline{0.00})                      & 13.45 (\underline{0.00})                        & 6.15 (\underline{0.00})                               \\
\textbf{Seasonal ARIMA (SARIMA)}                              & \textbf{10.45 (/)}                & \textbf{10.47 (/)}                & \textbf{10.48 (/)}                & \textbf{10.70 (/)}                  & 18.40 (\underline{0.00})                              \\
\textbf{Seasonal decomposition by Loess model (DM)}           & 11.80 (0.07)                      & 11.80 (0.08)                      & 11.81 (0.07)                      & 12.12 (\underline{0.03})                        & 9.67 (\underline{0.00})                               \\
\textbf{Seasonal random walk (SRW)}                           & 12.40 (\underline{0.00})                      & 12.43 (\underline{0.00})                      & 12.45 (\underline{0.00})                      & 13.14 (\underline{0.00})                        & 2.62 (1.00)                               \\
\textbf{Simple Multilayer Perceptron (MLPUni)}                & 12.52 (\underline{0.00})                      & 12.61 (\underline{0.00})                      & 12.61 (\underline{0.00})                      & 12.59 (\underline{0.00})                        & 21.95 (\underline{0.00})                              \\
\textbf{Support Vector Regression (SVRUni)}                   & 12.29 (\underline{0.00})                      & 12.29 (\underline{0.00})                      & 12.30 (\underline{0.00})                      & 12.45 (\underline{0.00})                        & 17.55 (\underline{0.00})                              \\ \midrule
\multicolumn{1}{l}{\textit{With external factors}}                     & \multicolumn{1}{l}{}              & \multicolumn{1}{l}{}              & \multicolumn{1}{l}{}              & \multicolumn{1}{l}{}                & \multicolumn{1}{l}{}                      \\ \midrule
\textbf{Conditional Inference Regression Tree (CtreeMulti)}   & 12.93 (\underline{0.00})                      & 12.93 (\underline{0.00})                      & 12.88 (\underline{0.00})                      & 12.81 (\underline{0.00})                        & 9.76 (\underline{0.00})                               \\
\textbf{K Nearest Neighbors Regression (KNNMulti)}            & 15.07 (\underline{0.00})                      & 15.07 (\underline{0.00})                      & 15.10 (\underline{0.00})                      & 14.90 (\underline{0.00})                        & 18.53 (\underline{0.00})                              \\
\textbf{Long Short Term Memory RNN (LSTMMulti)}               & 17.05 (\underline{0.00})                      & 16.93 (\underline{0.00})                      & 16.95 (\underline{0.00})                      & 16.78 (\underline{0.00})                        & 25.02 (\underline{0.00})                              \\
\textbf{Multiple Linear Regression (LRMulti)}                 & 12.89 (\underline{0.00})                      & 12.88 (\underline{0.00})                      & 12.86 (\underline{0.00})                      & 12.77 (\underline{0.00})                        & 6.05 (\underline{0.00})                               \\
\textbf{Multivariate Adaptive Regression Splines (MARSMulti)} & 13.25 (\underline{0.00})                      & 13.25 (\underline{0.00})                      & 13.24 (\underline{0.00})                      & 13.22 (\underline{0.00})                        & 11.87 (\underline{0.00})                              \\
\textbf{Random Forests (RFMulti)}                              & 14.14 (\underline{0.00})                      & 14.12 (\underline{0.00})                      & 14.10 (\underline{0.00})                      & 13.89 (\underline{0.00})                        & 15.61 (\underline{0.00})                              \\
\textbf{Recursive Partitioning Regression Tree (RpartMulti)}  & 14.28 (\underline{0.00})                      & 14.31 (\underline{0.00})                      & 14.29 (\underline{0.00})                      & 13.98 (\underline{0.00})                        & 8.70 (\underline{0.00})                               \\
\textbf{Seasonal ARIMAX (SARIMAX)}                            & 10.81 (1.00)                      & 10.80 (1.00)                      & 10.82 (1.00)                      & 11.16 (\underline{0.00})                        & 20.78 (\underline{0.00})                              \\
\textbf{Simple Multilayer Perceptron (MLPMulti)}              & 14.13 (\underline{0.00})                      & 14.11 (\underline{0.00})                      & 14.12 (\underline{0.00})                      & 13.45 (\underline{0.00})                        & 24.20 (\underline{0.00})                              \\
\textbf{Support Vector Regression (SVRMulti)}                 & 14.04 (\underline{0.00})                      & 14.01 (\underline{0.00})                      & 14.04 (\underline{0.00})                      & 14.06 (\underline{0.00})                        & 19.13 (\underline{0.00})                              \\
\textbf{Vector Autoregression (VAR)}                          & 13.97 (\underline{0.00})                      & 13.99 (\underline{0.00})                      & 14.00 (\underline{0.00})                      & 14.05 (\underline{0.00})                        & 14.31 (\underline{0.00})                              \\ \midrule
\multicolumn{1}{l}{\textit{Friedman test}}                    & \multicolumn{1}{l}{}              & \multicolumn{1}{l}{}              & \multicolumn{1}{l}{}              & \multicolumn{1}{l}{}                & \multicolumn{1}{l}{}                      \\ \midrule
\textbf{Chi-Squared}                                          & 1299.8                            & 1285.8                            & 1286.7                            & 1059.3                              & 18708                                     \\
\textbf{P-value}                                              & \textless 2.2e-16                 & \textless 2.2e-16                 & \textless 2.2e-16                 & \textless 2.2e-16                   & \textless 2.2e-16                         \\ \bottomrule
\end{tabular}%
}
\caption{Overview of benchmarking results. Columns contain the forecasting techniques and their average ranks according to MAPE, RMSE, MASE, expected profit and computation time. The numbers between brackets are the p-values from the pairwise Nemenyi test that compares the given method to the best technique according to the evaluation metric at hand.}
\label{tab:res}
\end{table}

The results of the experiments are based on a total of 21840 forecasts, as we performed 24 one-month-ahead forecasts on 35 time series with 26 different models. We only take into account the results for models that have completed both the parameter tuning and feature selection procedures that were explained in the methodology section of this paper, see Section \ref{meth}. Other model set-ups were disregarded for the final analyses.

In order to compare all of these models to one another, we apply two ranking tests for the 26 forecasting techniques according to five evaluation measures: MAPE, RMSE, MASE, Profit and computation time. Concretely, we rank all of the methods for each of the 840 unique forecasts and then display the average over these forecasts. This methodology ensures a fairer comparison between the techniques than, e.g., simply taking an average MAPE of the 840 forecasts. Furthermore, we can verify if the differences in rank are significantly separate from one another. The Friedman test \citep{f4} is a non-parametric statistical test that verifies whether the difference between two treatments is significant or not. In this benchmark, the 26 forecasting techniques constitute the 'treatments', defined as $k$ in Equation \ref{eq9}, while the 35 time series datasets are the 'blocks', $N$ in Equation \ref{eq9}, which form groups of similar units. The Friedman test will rank the treatments according to a given evaluation criterion and will compare this ranking for each block. Therefore, three different average rankings are made for these experiments, according to the three evaluation measures. The p-value of the Friedman test then indicates if there exists a significant difference between any of the treatments.

\bigskip
\noindent
\begin{equation}
\label{eq9}
\chi_F^2=\frac{12N}{k(k+1)}[\displaystyle\sum_{j}R_j^2- \frac{k(k+1)^2}{4}]
\end{equation}
\indent
\bigskip

If this test is significant, a post-hoc analysis must follow, as we are interested to know which techniques differ from one another. As all of the Friedman tests are indeed significant, we turn to a second step, which consists of a pairwise Nemenyi test \citep{n1} for the three rankings. Concretely, the test determines if the average ranks of the models are at least at a critical distance of: 

\bigskip
\noindent
\begin{equation}
\label{eq10}
CD=q_{\alpha}\sqrt{\frac{k(k+1)}{6N}}
\end{equation}
\bigskip

\noindent
with $q_{\alpha}$ as critical values, which consist of the Studentized range statistic divided by $\sqrt{2}$.

\begin{figure}[h]
    \centering
\includegraphics[width=13cm, trim = 0mm 12mm 5mm 5mm, clip]{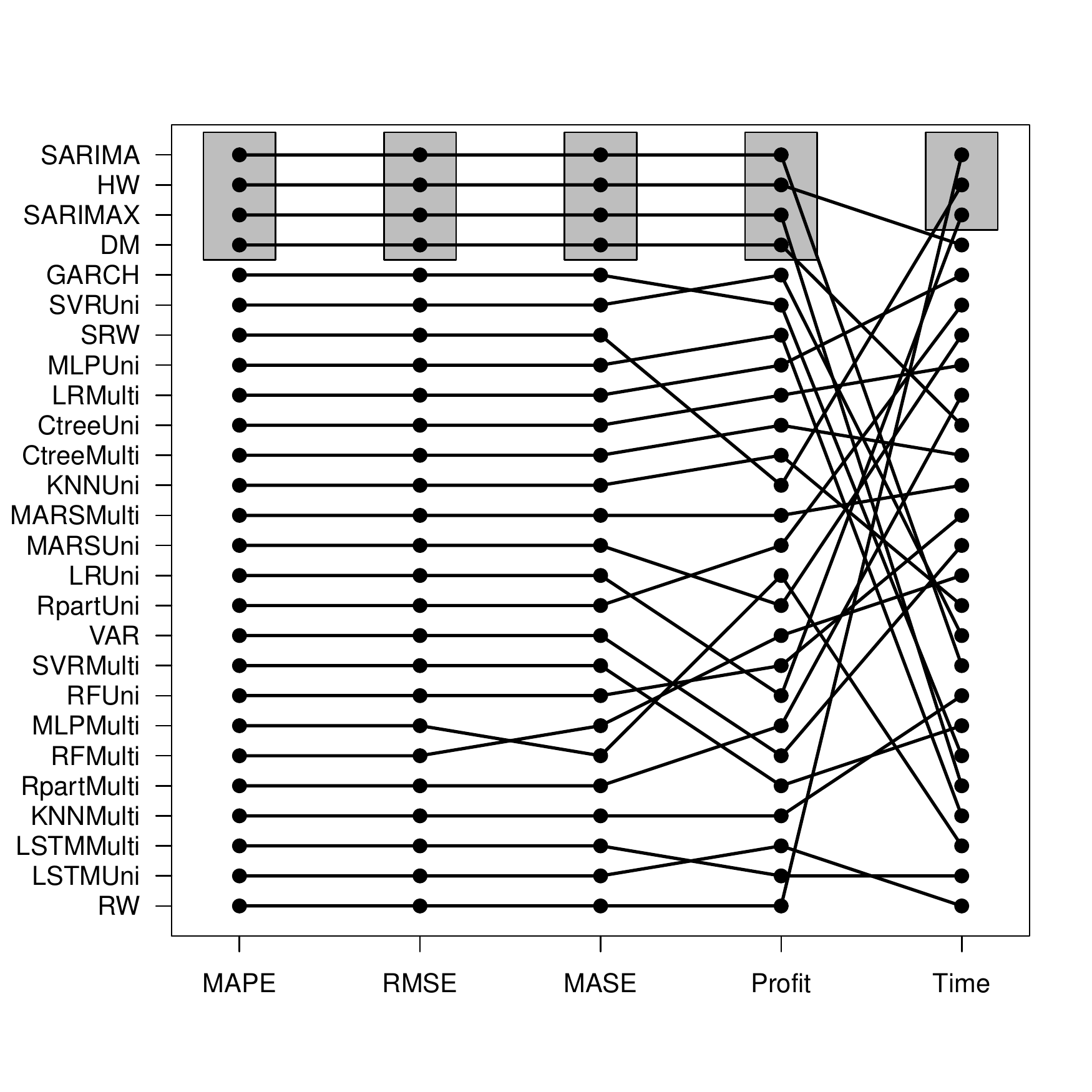}
 \caption{Rank comparison}
    \label{fig:rank}
\end{figure}

The results of these tests can be found in Table \ref{tab:res}. The first column contains the method’s name, together with its abbreviation in all following figures. This table then displays the average rank of all of the forecasting methods according to MAPE, RMSE, MASE, expected profit and computation time. The numbers between brackets are the p-values from the pairwise Nemenyi test that compares the given method to the best technique according to the evaluation metric at hand, which is indicated in bold for each evaluation measure. All p-values with a $95\%$ significance level are underlined. Figure \ref{fig:rank} displays the relative rankings according to each evaluation metric from best at the top to worst on the bottom. The grey boxes contain the models that are not significantly different from the best model at a $95\%$ significance level. The lines connect the relative rankings of the same technique according to the different measures.

\begin{figure}[h]
  \centering
  \includegraphics[width=7cm, clip]{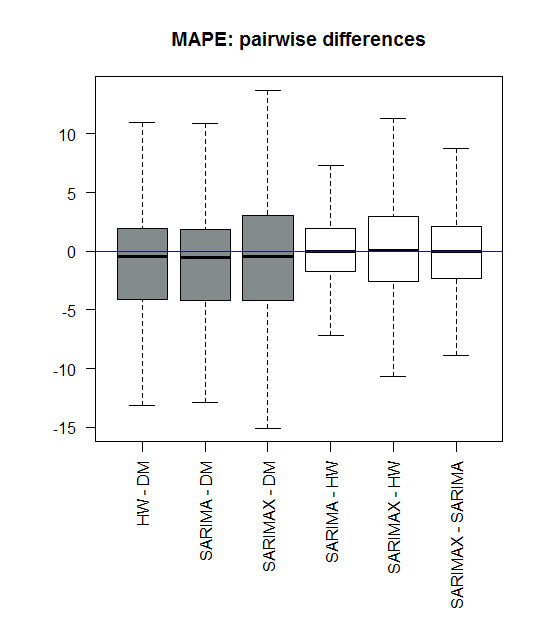}
  \includegraphics[width=7cm, clip]{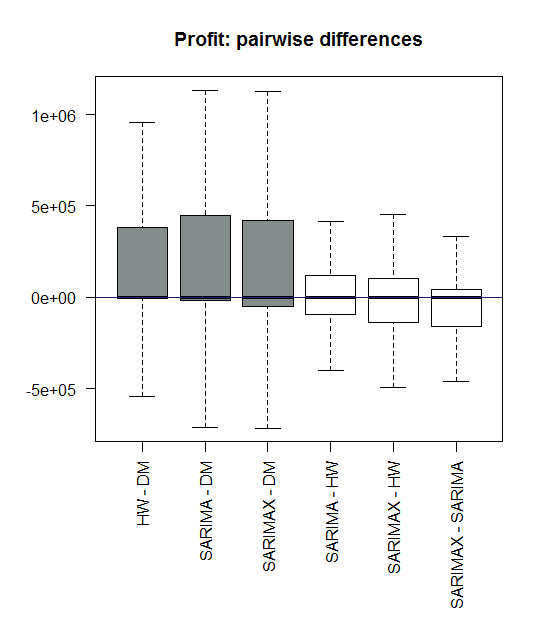}
 \caption{Pairwise differences of four best-performing models}
  \label{fig:pair}
\end{figure}

We can now turn back to our research questions, which are firstly focused on selecting the best type of technique for a tactical sales forecast. In our introduction, we posed three sub research questions that each require the techniques to be divided into two categories. Firstly, the techniques can be divided into techniques with and without external features. Furthermore, we contrast seasonal and non-seasonal methods, and distinguish between machine learning and statistical techniques. In order to ascertain whether one category significantly outperforms the other, we perform a Wilcoxon signed rank test on the average ranks of each category for each forecast. The null hypothesis of this test states that there is no difference between the two groups. For each of the categorizations, this test rejected the null hypothesis and proved to be significant at the $99\%$ significance level. Firstly, univariate techniques without variables outperform the ones that add external factors with a p-value smaller than 2.2e-16. Seasonal models have a significantly lower average rank, with a p-value that is also smaller than 2.2e-16. Finally, the ML techniques consistently lead to worse results than the statistical models, as the Wilcoxon test was highly significant with a p-value of 0.00235. In short, time series models that explicitly model seasonality and do not incorporate external factors still seem to be the best candidate for our set of tactical sales forecasts, which all display a trend and seasonality. In order to select the top performing techniques, we take Table \ref{tab:res} and Figure \ref{fig:pair} into consideration again, which show a clear top four models in terms of both accuracy and profit. The best-performing forecasting techniques are SARIMA, SARIMAX, Holt-Winters and the seasonal decomposition model (DM). The only exception to the seasonal, univariate and non-ML rule is SARIMAX, which also incorporates external drivers into the model.

In Figures \ref{fig:pair}, we take a closer look at these best-performing models in terms of both accuracy and the expected profit. These figures contain the distributions of the pairwise differences of SARIMA, SARIMAX, Holt-Winters and the seasonal decomposition model (DM). Grey boxplots indicate a significant difference between the two models that are mentioned on the Y-axis. DM seems to consistently perform worse than the other three models, while the remaining three time series models all perform equally in terms of both MAPE and expected profit.

The last performance metric that can still make a difference in the selection of the best-performing technique, is computation time. This measure is indicative of the complexity of the model, but can also have an effect on the final costs of the model as computation efforts also lead to additional expenses. The average computation time of the top four models is summarized in Table \ref{fig:comp} below. Clearly, the training of Holt-Winters and DM take the least amount of time by far. However, the average computation times of both SARIMA and SARIMAX are still below 10 seconds per forecast. Furthermore, it is logical that these last two techniques require more time to train, given the feature and hyper parameter optimization according to profit for both of them. In conclusion, a top three of equally performing time series models remains: Holt-Winters exponential smoothing, Seasonal ARIMA and Seasonal ARIMAX, but Holt-Winters will significantly save on computation time if there is a large number of time series to forecast.

\begin{table}[]
\centering
\begin{tabular}{@{}ll@{}}
\toprule
\textbf{Model}                        & \textbf{Average computation time (seconds)} \\ \midrule
\textbf{Holt-Winters}                 & 0.02                                        \\
\textbf{Seasonal decomposition model} & 0.05                                        \\
\textbf{SARIMA}                       & 3.89                                        \\
\textbf{SARIMAX}                      & 9.35                                        \\ \bottomrule
\end{tabular}
\caption{Average computation time for best-performing models}
\label{fig:comp}
\end{table}

\begin{figure}[h]
  \centering
  \includegraphics[width=13cm, trim = 0mm 30mm 0mm 0mm, clip]{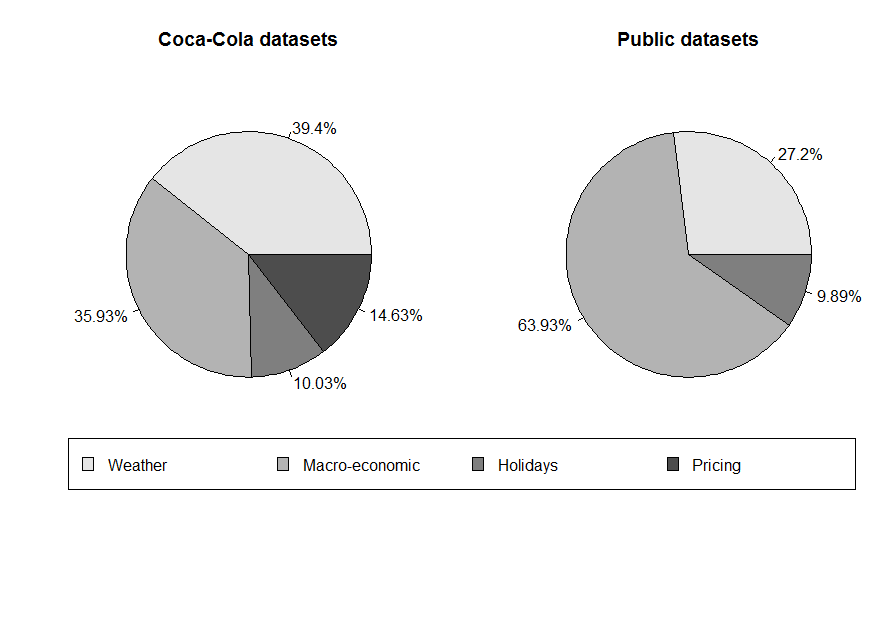}
 \caption{Average percentage of selected variables}
  \label{fig:selvar}
\end{figure}

Finally, we also take a closer look at the interpretability of these top three techniques. As time series models, they are all transparent methodologies that attribute weights to the autoregressive, trend and seasonal components of the time series. Additionally, SARIMAX displays the weights of the added external factors, indicating their impact on the sales, which greatly adds to the explanatory power of the model. This therefore entails a large advantage for the SARIMAX technique in terms of business value. On the other hand, the feature selection procedure leads to a higher computation time and effort, so these two aspects need to be weighed against one another. In the end, the univariate time series models perform equally to SARIMAX, but additional information on the external influences on the sales might be preferable in a business context. Note that this refers to two completely different objectives, i.e. predicting versus explaining. Before the final selection of the best technique, businesses need to clearly outline the objective of a forecasting model. In terms of variable selection in this paper, Figure \ref{fig:selvar} shows the average percentage of selected variables for each of the variable types, illustrated for each of the two data sources. From these charts, we can conclude that weather and macro-economic variables are selected the most for all datasets. On average, 2 weather variables and 2.5 macro-economic variables were selected for the Coca-Cola Company datasets, while 1.78 weather variables and 3.89 macro-economic variables were chosen for the public datasets.

\begin{figure}[h]
  \centering
  \includegraphics[width=13cm, clip]{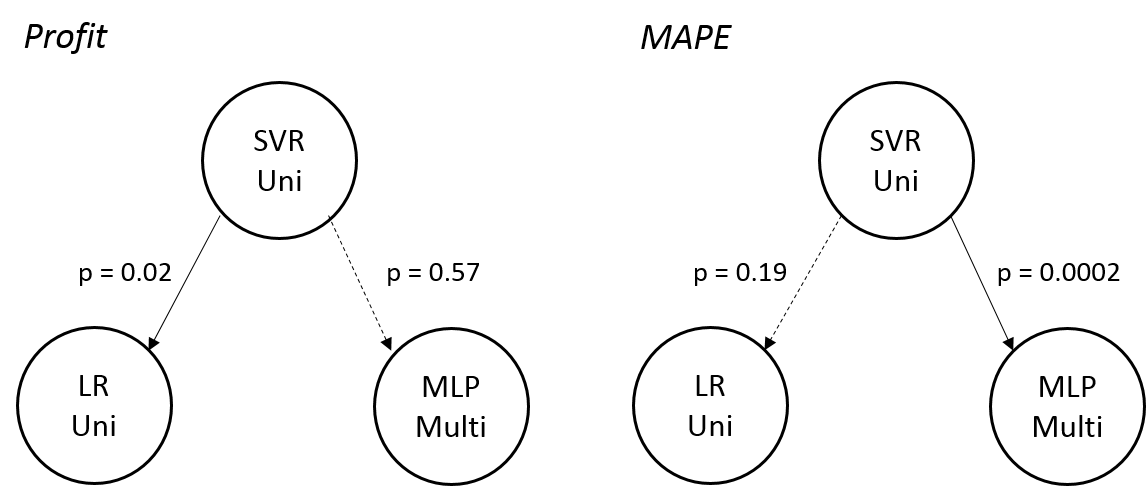}
 \caption{Domination plots of pairwise Nemenyi differences in p-values}
  \label{fig:dom}
\end{figure}

The second research question focused on the integration of the expected profit function into the model selection process. We can clearly see from Table \ref{tab:res} and especially Figure \ref{fig:rank} that the ranking of the techniques according to MAPE, RMSE and MASE are virtually the same, while the ranking according to the expected profit function looks a bit different. Although the top methods perform well according to all of these evaluation measures, the changes in the ranking already indicate that it is valuable to compare models according to profit as well, as it might lead to a different ranking of the possible techniques. For example, the p-values in Table \ref{tab:res} of the DM technique are not significantly different from the top three time series models in terms of the accuracy measures, but they are significantly different from them when we look at the expected profit function. In Figure \ref{fig:dom}, we will look at some pairwise differences of other models according to MAPE and Profit as well. In this figure, we can clearly see that techniques can significantly differ in terms of Profit and not in terms of MAPE, or vice versa. Specifically, we compare the univariate cases of Multiple Linear Regression (LRUni) and Support Vector Regression (SVRUni), and the variant with external factors of Simple Multilayer Perceptron (MLPMulti). The pairwise differences between SVR and LR, and SVR and MLP show that there is a clear difference between the two evaluation measures. It is also important to note that these changes do not exist in pairwise differences when we only compare the accuracy metrics.

\subsection{Discussion}
\label{Disc}

The results of this study have three interesting implications for model selection in sales forecasting from a business perspective. Firstly, we proposed a profit-driven approach that provides a completely automated framework for model building and selection. The expected profit function that we implement, is completely adaptable to any sales forecasting situation by combining business expertise with traditional accuracy-based evaluation. Furthermore, this profit function can be used as an evaluation criterion that gives a different view on which technique is truly the best one in a benchmarking exercise. While the results in this paper are consistently in line with the accuracy measures, the overall ranking according to profit is still significantly different than the accuracy-based ones. This indicates that a ranking according to profit might yield a different result in model selection. In this paper, however, the top three models' performance was consistenly very close to one another, while the same models outperformed others by a significant margin. In other cases, when model performance between techniques is closer to one another, the expected profit function can provide an additional perspective into final model selection. Furthermore, this paper adds to the scarce literature on the use of profit-driven analytics in forecasting and regression analysis.

Secondly, we notice that univariate time series models that explicitly capture seasonality, perform the best in this benchmarking study, although the Seasonal ARIMAX method is an exception to the univariate characteristic. However, this technique only performs equally to the aforementioned univariate methods and we can therefore raise the question if the addition of external variables is truly useful in this context. While other studies have shown the value of adding external drivers into the models for sales forecasting on a strategic level \citep{s1}, this research shows that we can forecast as well without any independent variables in the model. When we take into account the additional cost of data collection and model maintenance, we conclude that forecasting the sales on a product category level is easier achieved by univariate models without compromising on accuracy or profit. Although we recognize the added explanatory value of integrating features, we question if it is worth the effort when achieving the best forecast is the goal.

Finally, we compared two categories of forecasting techniques to one another: statistical methods and machine learning techniques. In the case of tactical sales forecasting, we clearly see that simpler models outperform the others significantly for these 35 datasets. This leads us to conclude that the more traditional models are actually still performing the best when tackling this type of time series problem. These findings are in line with \citep{m2}, but contradict \citep{c3}. To conclude, seasonal time series models tend to outperform other techniques for a tactical sales forecast. From a business perspective, this conclusion is especially positive, as these models are easy to interpret and have a faster computation time.

\section{Conclusion}
\label{Conc}

In this paper, we introduced a new, completely automated and profit-oriented strategy to sales forecasting, which integrates an expected profit function into several steps of model selection. This function can be implemented in any sales forecasting context by letting business experts and previous data set the profit margins for every product. Furthermore, our research has proven that simpler time series models tend to outperform more complex techniques for 35 sales datasets. All of the applied ML techniques achieve significantly worse results than the traditional models, both in accuracy and profit. This implies that less complex techniques are still the best type of method to handle tactical sales forecasting. Finally, we found that univariate time series models that are able to explicitly model the seasonality of a time series, perform best. This indicates that the addition of external variables is unnecessary, especially when we consider the additional costs that are linked to maintaining models with external drivers.

In terms of possible limitations of this study, we recognize some shortcomings in this paper. Firstly, it is impossible to come up with an exhaustive list of forecasting techniques. However, we have attempted to implement common methods from all three categories of techniques that are frequently used for forecasting. Furthermore, this research consists of 35 monthly time series, which is significantly less than the larger benchmarks and competitions in the field \citep{a6, c3, m2}. However, this paper particularly focuses on one field, i.e. sales forecasting, and is one of the larger benchmarking studies in this specific area. Furthermore, we have added to the generalizability and reproducibility of the study by including several publicly available datasets as well. Finally, we only implemented individual forecasting methods without considering ensemble methods. This type of methodology has become extremely popular in forecasting \citep{l1} and it has been proven that this approach can significantly impact the accuracy of forecasts. Potential future research therefore includes an expansion of this study in three aspects. Firstly, we can include more sales time series in order to further underpin our statements. Secondly, we can implement more techniques and include ensemble methods. Finally, this study can be further expanded to other fields than sales forecasting as well. However, given the range of techniques and the number of datasets that were already used in this paper, we can state that simple, seasonal time series models are still the best choice for a high-level tactical sales forecast.

\section*{Acknowledgements}
\label{ack}
We would like to acknowledge The Coca-Cola Company for funding this research and providing us with the necessary business expertise and data to conduct our experiments.

\section*{Bibliography}
\label{bib}
\bibliographystyle{elsarticle-harv} 
\bibliography{biblio}
\end{document}